\documentclass{PoS}
\usepackage{amsmath,amssymb}

\title{Dark radiation: 21cm signals and laboratory tests}

\ShortTitle{Dark radiation: 21cm signals and laboratory tests}

\author{\speaker{Josef Pradler}\thanks{The talk reports on two
    separate papers \cite{Cui:2017ytb} and \cite{Pospelov:2018kdh}.}\\
  Institute of High Energy Physics, Austrian Academy of Sciences, Nikolsdorfer Gasse 18, 1050 Vienna, Austria\\
  E-mail: \email{josef.pradler@oeaw.ac.at}}

\abstract{It is entirely possible that our Universe is filled with
  dark radiation, such as SM neutrinos or new physics states, that are
  sourced by the decay of dark matter with cosmologically long
  lifetime.
  If non-thermal neutrinos produced such way carry sufficient energy,
  they can leave a detectable imprint in experiments designed to
  search for very weakly interacting particles: dark matter and
  underground neutrino experiments.  Conversely, a very soft
  non-thermal population of dark photons sourced this way is allowed
  to exceed the number density of CMB photons by many orders of
  magnitude without being in conflict with current bounds. Equipped
  with a small probability of conversion into ordinary photons, the
  scenario becomes testable through the cosmological 21cm signal.}

\FullConference{Neutrino Oscillation Workshop (NOW2018)\\
		9 - 16 September, 2018\\
		Rosa Marina (Ostuni, Brindisi, Italy)}

\begin{document}

\section{Late Dark Radiation}

Observational cosmology over the past decade has resulted in a very
sensitive constraint on the number of extra degrees of freedoms that
remained radiation-like during the CMB epoch. Phrased in terms of
neutrino-like radiation species it reads as:
$N_{\rm eff} =3.04\pm 0.33$~\cite{Ade:2015xua} which translates into a
limit on the energy density in additional dark radiation (DR),
$\rho_{\rm DR}$, when measured relative to the one in photons,
$\rho_{\rm DR}/\rho_{\rm CMB} < 0.15 $.  This bound has a wide degree of
applicability, and is most effectively used to constrain models with
``early'' DR, or models with extra light degrees of freedom that were
in thermal contact with the SM, but decoupled at some point in the
history of the early Universe.  However, the constraint would not be
applicable to models where DR is created much later than the cosmic
microwave background (CMB) epoch. Indeed, late decays of a sizable
fraction of dark matter (DM) into dark radiation are allowed, and,
moreover, $\rho_{\rm DR}$ can be much larger than $\rho_{\rm CMB}$ today.

Two logical possibilities in the late generation of $\rho_{\rm DR}$
arise: either the number density of DR particles, $n_{\rm DR}$, is
smaller than that of CMB photons, $n_{\rm CMB}$, while their kinetic
energy, $E_{\rm DR}$, on average is much larger than the energy
$E_{\rm CMB}$ of individual CMB quanta,
\begin{align}
n_{\rm DR} \ll n_{\rm CMB},~~ E_{\rm DR} \gg
E_{\rm CMB},~~\rho_{\rm DR} (\sim E_{\rm DR}n_{\rm DR})\leq 0.1
\rho_{\rm DM}, 
\label{ineq}
\end{align}
or DR quanta are energetically much softer, but more numerous than CMB
photons,
\begin{equation}
  E_{\rm DR} \ll E_{\rm CMB},\quad  n_{\rm DR} > n_{\rm RJ}
  , \quad   \rho_{\rm DR} (\sim E_{\rm DR}n_{\rm DR})\leq 0.1 
\rho_{\rm DM}  . 
\label{setup}
\end{equation}
In the last relations of (\ref{ineq}) and (\ref{setup}) we require
that the amount of DR does not exceed 10\% of the DM energy density,
in accordance with recently updated constraints~\cite{Poulin:2016nat};
$n_{\rm RJ}$ represents the low-energy Rayleigh-Jeans (RJ) tail of the
standard CMB Planck distribution up to an energy $E_{\rm max}\ll T_{\rm CMB}$,
$n_{RJ} \approx T_{\rm CMB} E^2_{\rm max} / (2 \pi^2) $.
%
%with
%$n_{\rm CMB} = 2\zeta(3)/\pi^2\,T_{\rm CMB}^3 \simeq 0.24\,T_{\rm
%  CMB}^3$ being the full Planckian number density.
%
Such set of inequalities leaves, of course, a lot of freedom for what
DR can be, but restricts a number of possibilities for how the
non-thermal DR can be created. The most efficient mechanism for
populating DR radiation states is the decay of DM, and if it happens
at a redshift after CMB decoupling, a per-cent fraction of DM is
allowed to decay to DR.

There are then two principal components of the DR flux, either
originating from DM decay within the galaxy or from extragalactic
distances. The former is obtained from a standard line-of-sight
integral over the galactic DM distribution, the latter is obtained as
a redshift integral over the cosmological DM density.  Focusing on a
mono-chromatic injection of a pair of new particles $X$ where $X$ is either
assumed to be SM neutrino $X = \nu$ or $\bar \nu$ or a dark photon
$X=A'$, the cosmological energy spectrum at redshift $z$ is given by
\begin{align}
  \label{eq:DPspec}
  \frac{dn_{X} (E, z)}{dE} =
  \frac{2 \Omega_{\rm dm} \rho_c (1+z)^3  }{m_{\rm dm} \tau_{\rm dm} E H(\alpha-1)}
  \Theta(\alpha - 1 - z).
\end{align}
Here, limits of cosmologically long DM lifetime
$\tau_{\rm dm} > t_0^{-1}$ and of relativistic injection energy
$E_X\gg m_{X}$ have been taken. Furthermore, $\rho_c $ is the critical
density, $\Omega_{\rm dm} h^2 = 0.12$ the DM density
parameter~\cite{Ade:2015xua}, and the Hubble rate, $H(z)$, is
evaluated at redshift $\alpha-1$, where
$\alpha \equiv m_{\rm dm} (1+z)/(2E)$; $H_0 = H(t_0)$ and $t_0$ is the
current age of the Universe.

\section{Energetic Neutrino Dark Radiation: Laboratory Tests}
\label{sec:energ-neutr-dark}

In this scenario~\cite{Cui:2017ytb} DM gives rise to DR in the form of
SM neutrinos with initial energies of several tens of~MeV. This can
occur directly by DM decay to $ \nu(\bar \nu)+Y$, where $Y$
stands for the rest of the decay products, or in two steps: first
through the decay of DM to a nearly massless fermion ({\em e.g.}, a
sterile neutrino), that then oscillates into the SM neutrinos.  Models
with direct decay to neutrinos are free from potential constraints
from $N_{\rm eff}$ measurements, the $\nu$-SM interactions are known,
and the model is more minimal/simple. Then, on top of the constraints
from dark matter experiments that limit the neutral current
interactions of the DR neutrinos with nuclei, there will be additional
constraints provided through weak interactions at neutrino detectors
such as Super-Kamiokande (SK).

The total flux $\phi_{\rm tot}$, integrated over the whole energy
spectrum, varies over many orders of magnitude depending on the choice
parameters. Nevertheless, one can estimate the maximum possible flux
at $\tau_{\rm dm} = 100\, \rm Gyr$, while taking $m_X \to 0$, and
keeping $m_{\rm dm}$ as a free parameter:
$\phi_{\rm tot}^{\rm max} \sim ({10\,{\rm MeV}}/{m_{\rm dm}} )\times
10^6 \, {\rm cm}^{-2} {\rm s}^{-1}. $
Coincidentally, the value of the DR flux may become comparable to that
of $^{8}$B solar neutrinos at $m_{\rm dm} \sim 10$\,MeV, and exceeds the diffuse
SN neutrino flux by many orders of magnitude at $m_{\rm dm} \sim 50$\,MeV.
With regards to the injection of SM neutrinos (and not
anti-neutrinos), it turns out that constraints from SK are superior to
the current sensitivity of DM direct detection experiments. To pick a
specific example, the current constraint on the neutrino flux
originating from the decay of a DM particle of $m_{\rm dm}=50$\,MeV and
lifetime $\tau_{\rm dm} \gg t_0$ :
$  \phi_{\nu}(E_\nu \simeq 25\,{\rm MeV}) < 5\times 10^{2} {\rm
    \,cm}^{-2}{\rm s}^{-1}. $
This constraint is more than two orders of magnitude more relaxed
compared to the SK limit on a cosmic $\bar\nu_e$ flux.  Consequently,
if this limit is saturated by DR, then the expected scattering rate
inside the xenon-based direct detection experiments will be such as to
mimic a DM particle with $m_{\rm dm} \sim 30$\,GeV and cross section
of $\sigma \sim 10^{-47}$cm$^2$, which is significantly above the
traditionally derived ``neutrino floor''.

\section{Soft dark photon radiation: 21cm signals}
\label{sec:soft-dark-photon}

Returning to (\ref{setup}) and considering the case when typical DR
energy is significantly smaller than that of CMB
photons~\cite{Pospelov:2018kdh}, one observes that the number of DR
quanta may significantly exceed $n_{\rm RJ}$.  For example, letting
$5\%$ of the DM energy density convert to DR in the frequency range
$E_{\rm max}/T_{\rm CMB } =2\times 10^{-3}$ after CMB decoupling,
$n_{\rm DR} \leq 3.3 \times 10^{5} \, n_{\rm CMB}$ where
$n_{\rm CMB} = 2\zeta(3)/\pi^2\,T_{\rm CMB}^3 $ is the full Planckian
number density. Thus, soft DR quanta can outnumber the RJ CMB photons
by up to 11 orders of magnitude.

For the case for SM neutrinos considered above that have Fermi-type
interactions with atomic constituents, this type of DR would be very
difficult to see directly.  There is, however, one class of new fields
that can manifest their interactions at low energies and low
densities.  These are light vector particles (often called dark
photons), $A'$, that develop a mixing with ordinary photons,
$\epsilon F'_{\mu\nu}F_{\mu\nu}$.  The apparent
number count of CMB radiation can be modified by photon/dark photon
oscillation:
\begin{equation}
\frac{dn_{A}}{dE} \to \frac{dn_{A}}{dE} \times P_{A\to A} + \frac{dn_{A'}}{dE} \times P_{A'\to A}~,
\label{modified}
\end{equation}
where $P_{A\to A} = 1 - P_{A\to A'}$ is the photon survival
probability, while $ P_{A'\to A}$ is the probability of $A'\to A$
conversion.  The physics of the 21\,cm line then provides a useful
tool to probe DR through the apparent modification of the low-energy
tail of the CMB:
at the end of the dark-ages, the strength of the net-absorption signal
of 21~cm radiation through the hyperfine transition in neutral
hydrogen is expected to be proportional to $1-T_{\rm CMB}/T_{s}$. Here
$T_{\rm CMB}$ is a proxy that counts the number of CMB photons
interacting with the two-level hydrogen hyperfine system, and $T_s$ is
the spin temperature.  The relevant photons have energy
$E_0 = 5.9~\mu\mathrm{eV}$ at redshift $z \approx 17$, and
therefore reside deep within the RJ tail,
$E_0/T_{\rm CMB} \approx 1.4 \times 10^{-3}$. Cosmological $A\leftrightarrow A'$
oscillations that populate photons in the relevant frequency band may
therefore enhance the strength of the absorption signal.

Indeed, the EDGES experiment recently presented a tentative detection
of 21\,cm absorption coming from the interval of redshifts
$z=15-20$~\cite{Bowman:2018yin}, indicating a twice as strong 
absorption signal than expected.
We find that the oscillation of non-thermal DR into visible photons
can easily accommodate EDGES consistent with other constraints. For
this a resonance condition in the $A' \to A$ conversion must be met, $m_{A'}=m_{A}(z)$, where
$m_A(z)$ is the plasma mass of photons at redshift~$z$. In the course
of cosmological evolution
$m_{A}(z) \simeq 1.7 \times 10^{-14}{\rm eV} \times (1+z)^{3/2}
X^{1/2}_e(z) $ scans many orders of magnitude where $X_e$ is the free
electron fraction, and hence allows a sizable permissible dark photon mass
range.
For example, for DM decay with $m_{\rm dm} = 10^{-3}$~eV,
$\tau_{\rm dm} = 100~t_0$ and a dark photon mass of
$m_{A'} = 5\times 10^{-12}\, \rm eV$ (implying a resonance redshift
$z_{\rm res} = 500$), and $\epsilon = 2.1\times 10^{-7}$, the number
of photons in the relevant frequency interval is doubled at
$z=17$---implying a twice as strong 21~cm absorption signal---while the
total energy density of dark photons relative to the energy density of
ordinary CMB photons at resonance is $6 \times 10^{-6}$, and a
fraction $4 \times 10^{-4}$ of the dark photons oscillate into
ordinary photons.

\vspace*{-0.1cm}
\section{Conclusions}
\label{sec:conclusions}

A cosmic DR background can be probed through a variety of
observational and laboratory tests.  SM neutrinos with energy
$O(10)$~MeV are tested in direct detection and neutrino
experiments~\cite{Cui:2017ytb}. Dark photons with energies well below $T_{\rm CMB}$ are tested through  21~cm cosmology~\cite{Pospelov:2018kdh}.\medskip \\
\textit{Acknowledgements:} JP acknowledges the collaboration with
Y.~Cui, M.~Pospelov, J.~Ruderman and A.~Urbano on the presented
papers.  The work is supported by the New Frontiers Program by the Austrian
Academy of Sciences.

\bibliographystyle{JHEP}
\bibliography{nowRefs}

\end{document}